\documentclass[a4paper,12pt]{article}
\usepackage{multirow}
\usepackage{verbatim}
\usepackage{amsfonts}
\usepackage{mathrsfs}
\usepackage{amsmath}
\usepackage{amssymb}
\usepackage{float}
\usepackage{framed}
\usepackage[medium]{titlesec}
\usepackage{bm}
\usepackage{cite}
\usepackage[normalem]{ulem}
\usepackage{extarrows}
\usepackage{slashed}
\usepackage{isodateo}
\usepackage{graphicx}
\usepackage{xcolor}
\usepackage[bookmarksnumbered=true,bookmarksopen=true,hyperfootnotes=true]{hyperref}
\hypersetup{colorlinks,%
	linkcolor=[rgb]{0,0.3,0.6}, %
	citecolor=[rgb]{0,0.3,0.6}, %
	urlcolor=[rgb]{0,0.3,0.6}}
\usepackage[hmargin=.7in,vmargin=1.1in]{geometry}
\usepackage{indentfirst}
\usepackage{booktabs}

\usepackage{enumerate}
\usepackage{amsfonts}
\usepackage{mathrsfs}
\usepackage{amsmath}
\usepackage{amssymb}
\usepackage{float}
\usepackage{bm}
\usepackage{slashed}
\usepackage{feynmp-auto}
\usepackage[normalem]{ulem}
\usepackage{extarrows}
\usepackage{slashed}
\usepackage{isodateo}
\usepackage{graphicx}
\usepackage{xcolor}
\usepackage{cancel}

\definecolor{bblue}{rgb}{0,0.2,0.6}

\newcommand{\fr}[2]{\mbox{$\frac{\,{#1}\,}{#2}$}}

\renewcommand{\rm}{\mathrm}

\usepackage{indentfirst}

\graphicspath{{fig/}}

\usepackage{tikz}
\usetikzlibrary{arrows,shapes}
\usetikzlibrary{trees}
\usetikzlibrary{matrix,arrows} 				
\usetikzlibrary{positioning}				
\usetikzlibrary{calc,through}				
\usetikzlibrary{decorations.pathreplacing}  
\usepackage{pgffor}							

\usetikzlibrary{decorations.pathmorphing}	
\usetikzlibrary{decorations.markings}
\tikzset{
	photon/.style={decorate, decoration={snake}, draw=red},
	electron/.style={draw=blue, postaction={decorate},
		decoration={markings,mark=at position .55 with {\arrow[draw=blue]{>}}}},
	gluon/.style={decorate, draw=black,
		decoration={coil,amplitude=4pt, segment length=4pt}} ,
	vector/.style={decorate, decoration={snake}, draw},
	provector/.style={decorate, decoration={snake,amplitude=2.5pt}, draw},
	antivector/.style={decorate, decoration={snake,amplitude=-2.5pt}, draw},
	fermion/.style={draw=black, postaction={decorate},
		decoration={markings,mark=at position .55 with {\arrow[draw=black]{>}}}},
	fermionbar/.style={draw=black, postaction={decorate},
		decoration={markings,mark=at position .55 with {\arrow[draw=black]{<}}}},
	fermionnoarrow/.style={draw=black},
	fermionnoarrowsoft/.style={draw=blue},
	scalar/.style={dashed,draw=black, postaction={decorate},
		decoration={markings,mark=at position .55 with {\arrow[draw=black]{>}}}},
	scalarbar/.style={dashed,draw=black, postaction={decorate},
		decoration={markings,mark=at position .55 with {\arrow[draw=black]{<}}}},
	scalarnoarrow/.style={dashed,draw=black},
	scalarnoarrowsoft/.style={dashed,draw=blue},
	electron/.style={draw=black, postaction={decorate},
		decoration={markings,mark=at position .55 with {\arrow[draw=black]{>}}}},
	bigvector/.style={decorate, decoration={snake,amplitude=4pt}, draw},
}

\tikzstyle{block} = [draw, rectangle, 
minimum height=3em, minimum width=6em]
\usetikzlibrary{arrows,patterns}

\def\mb{\mathbf}
\def\fr{\frac}

\newcommand{\email}[1]{\footnote{Email: \href{mailto:#1}{\nolinkurl{#1}}}}
\def\bgm{\begin{matrix}}
\def\edm{\end{matrix}}
\def\a{\mathsf{a}}
\def\b{\mathsf{b}}
\def\ma{\mathcal}
\def\la{\langle}
\def\ra{\rangle}
\def\fr{\frac}
\def\mb{\mathbf}

\newcommand{\be}{\begin{equation}}
\newcommand{\ee}{\end{equation}}
\newcommand{\ba}{\begin{array}}
\newcommand{\ea}{\end{array}}
\newcommand{\bea}{\begin{eqnarray}}
\newcommand{\eea}{\end{eqnarray}}

\renewcommand{\d}{\mathrm{d}}

\newcommand{\besub}{\begin{subequations}}
\newcommand{\eesub}{\end{subequations}}

\newcommand{\tchi}{{\tilde\chi}}

\newcommand{\beq}{\begin{equation} \begin{aligned}}
\newcommand{\eeq}{\end{aligned} \end{equation}}

\usepackage{mdframed}

\newmdenv[skipabove=0pt,%
skipbelow=5pt,%
leftmargin=0pt,%
rightmargin=0pt,%
innertopmargin=-5pt,%
innerbottommargin=7pt,%
innerleftmargin=2pt,%
innerrightmargin=2pt,%
splittopskip=0pt,%
splitbottomskip=0pt,%
linewidth=0pt,%
nobreak=true]%
{keyeqn2}

\newmdenv[backgroundcolor=gray!15,%
skipabove=0pt,%
skipbelow=5pt,%
leftmargin=0pt,%
rightmargin=0pt,%
innertopmargin=-5pt,%
innerbottommargin=7pt,%
innerleftmargin=2pt,%
innerrightmargin=2pt,%
splittopskip=0pt,%
splitbottomskip=0pt,%
linewidth=0pt,%
nobreak=true]%
{keyeqn}

\usepackage{titlesec} 
\begin{document}
\title{\Large\textbf{A model for inflaton induced baryogenesis and its phenomenological consequences }\\[2mm]}
\author{Haipeng An$^{1,2,\,}$\email{anhp@mail.tsinghua.edu.cn},~~ Qi Chen$^{1,\,}$\email{chenq20@mails.tsinghua.edu.cn}~~ and ~~Yuan Yin$^{1,\,}$\email{yiny18@mails.tsinghua.edu.cn}\\[5mm]
\normalsize{${}^{1}\,$\emph{Department of Physics, Tsinghua University, Beijing 100084, China}}\\ 
\normalsize{${}^{2}\,$\emph{Center for High Energy Physics, Tsinghua University, Beijing 100084, China}}\\ 
}
\date{}
\vspace{20mm}
\maketitle
\begin{abstract}
In this study, we introduce a novel approach aimed at addressing the longstanding baryon-anti-baryon asymmetry conundrum. Our proposed mechanism suggests that baryon numbers were generated during the inflationary epoch through the dynamics of the inflaton field coupled with an explicit baryon number violating interaction. Notably, during inflation, it is possible to halt the baryon number generation process via a symmetry restoration phase transition. We elucidate that prior to this phase transition, baryon numbers could be synthesized and preserved within classical field configurations. Subsequently, following the phase transition, these baryon numbers were liberated as particles. Crucially, we demonstrate that this mechanism of baryon number production is intricately linked with significant cosmological collider signals and gravitational wave (GW) signals, offering a compelling framework to explore the origins of the universe's matter-antimatter asymmetry.

\vspace{10mm}	
\end{abstract}
\newpage
\tableofcontents
\newpage

\section{Introduction}

The puzzle of baryon-anti-baryon asymmetry remains one of the most enigmatic and profound questions in the realm of cosmology and particle physics. 
To generate the asymmetry, the Sakharov conditions \cite{Sakharov:1967dj} must be satisfied. These pivotal conditions include: (1) the non-conservation of baryon number; (2) the violation of charge conjugation symmetry ($C$) and time-reversal symmetry ($T$); and (3) the necessity for the baryon number violating processes to occur out of thermal equilibrium. These conditions can be met through various mechanisms, including first-order phase transitions with baryon number and $CP$-violating interactions, as seen in electroweak baryogenesis~\cite{Kuzmin:1985mm}, or through the late decays of heavy Majorana particles with $CP$-violating interactions, exemplified by the leptogenesis model~\cite{Minkowski:1977sc,Fukugita:1986hr}. Additionally, the Affleck-Dine mechanism posits that baryogenesis may originate from the initial conditions of scalar quarks or leptons~\cite{Affleck:1984fy, Hertzberg:2013mba, Lozanov:2014zfa, Yamada:2015xyr, Bamba:2016vjs, Bamba:2018bwl, Cline:2019fxx, Barrie:2020hiu, Lin:2020lmr, Kawasaki:2020xyf, Kusenko:2014lra, Wu:2019ohx, Charng:2008ke, Ferreira:2017ynu, Rodrigues:2020dod, Lee:2020yaj, Enomoto:2020lpf,Wu:2021mwy}. Recent years have seen the emergence of innovative paradigms for baryogenesis, such as axiogenesis~\cite{Co:2019wyp,Co:2020jtv,Co:2021lkc,Co:2021qgl,Co:2022aav,Co:2022kul} and mesogenesis~\cite{Elor:2018twp,Alonso-Alvarez:2019fym,Elor:2020tkc,Elahi:2021jia}, which offer fresh perspectives on this longstanding issue. A detailed exploration of these recent advancements is available in comprehensive reviews~ \cite{Elor:2022hpa}.

The concept of an inflationary era preceding the thermal Big Bang of the universe has revolutionized our understanding of the cosmos, addressing the causality and flatness problems associated with the thermal Big Bang theory~\cite{Starobinsky:1979ty,Guth:1980zm,Linde:1981mu,Albrecht:1982wi}. Moreover, inflation provides a straightforward quantum mechanical mechanism for the genesis of the universe's large-scale structure as observed today~\cite{Mukhanov:1981xt,Hawking:1982cz,Starobinsky:1982ee,Guth:1982ec,Bardeen:1983qw}. It has also been suggested that the particles constituting dark matter today could have been generated during the inflationary period. Building on this foundation, our work introduces a novel scenario wherein the baryon-anti-baryon asymmetry is also produced during the inflationary epoch. This approach not only enriches our understanding of the early universe but also integrates the origins of matter asymmetry with the broader framework of cosmic inflation, offering a cohesive narrative for the genesis of the observable universe.

In our scenario, we assume the slow-roll inflation model. The rolling of the inflaton field plays an essential role in generating the baryon numbers. Unlike the Affleck-Dine mechanism, our model does not attribute baryon or lepton numbers to the inflaton field itself. Instead, within our theoretical construct, the inflaton is represented by a real scalar field $\phi$, while a complex scalar field $\chi$ is introduced to carry a $U(1)$ charge. This charge is pivotal, as it will subsequently be identified with either baryon or lepton number, laying the groundwork for our mechanism of baryogenesis or leptogenesis. 

The key mechanism at work involves the rolling of the homogeneous component of the inflaton field, denoted as $\dot\phi_0$, which effectively acts as a {\it chemical potential} for the $U(1)$ charge~\cite{Cohen:1988kt,Cohen:1991iu,Dine:1990fj,Chen:2018xck,Wang:2019gbi,Wang:2020ioa,Bodas:2020yho,Cui:2021iie,Tong:2022cdz}. This chemical potential, as we will demonstrate, plays a dual role. Firstly, in conjunction with explicit $U(1)$ symmetry-violating interactions, it leads to the production of a net $U(1)$ number, setting the stage for the asymmetry necessary for baryogenesis or leptogenesis. 
Secondly, and equally significant, this chemical potential, when paired with the explicit $U(1)$ breaking interaction, is capable of generating substantial signals detectable by cosmological collider experiments. These signals, as highlighted in~\cite{Bodas:2020yho}, offer a direct window into the high-energy processes occurring during the inflationary epoch. Consequently, the detection of such cosmological collider signals could serve as a definitive ``smoking gun" evidence for our model, providing not only validation for the proposed mechanism of baryon or lepton number generation but also offering invaluable insights into the physics of the early universe.
Through this innovative approach, our model bridges the gap between the microcosmic processes governing particle interactions and the macrocosmic observations of cosmological phenomena, offering a compelling framework for understanding the origins of matter asymmetry in the universe.

To explicitly break the $U(1)$ symmetry within our model, we introduce a real scalar field, $\sigma$, into the Lagrangian. Additionally, we posit the existence of a ${\cal Z}_2$ symmetry specifically within the interactions between the $\chi$ and $\sigma$ fields. The precise charge assignments of these fields are meticulously outlined in Table~\ref{tab:1}, where it is evident that the complex scalar $\chi$ uniquely carries the $U(1)$ charge. To facilitate the explicit breaking of the $U(1)$ symmetry, we incorporate a soft breaking term:
\begin{equation}\label{eq:Aterm}
A \sigma^2 \chi + {\rm {h.c.}}
\end{equation}
This term not only breaks the $U(1)$ symmetry but also introduces a tadpole term for $\chi$ upon the breaking of the ${\cal Z}_2$ symmetry. The presence of this tadpole term is instrumental in generating both the $U(1)$ number and the cosmological collider signal. However, a significant challenge arises as this term also tends to severely diminish the $U(1)$ number following inflation, a phenomenon known as the washout effect.
To mitigate this washout effect, we introduce a novel aspect to our model: the evolution of the inflaton field is assumed to trigger a phase transition that temporarily restores the ${\cal Z}_2$ symmetry at a certain juncture during inflation. Consequently, after this phase transition, the expectation value of $\sigma$ becomes zero, $\langle \sigma \rangle = 0$, rendering the washout effect perturbative and thus significantly less impactful.
In this work, we meticulously explore the parameter space of our model and identify regions where the washout effect post-phase transition is effectively negligible. Simultaneously, in these regions, a substantial cosmological collider signal is generated, offering promising avenues for empirical verification. Through this delicate balancing act, our model not only addresses the challenge of $U(1)$ number conservation post-inflation but also leverages the cosmological collider framework to provide tangible evidence for the underlying physics of baryogenesis during the inflationary epoch.

The occurrence of the phase transition within our model is not only pivotal for the dynamics of baryogenesis but also serves as a potential source of GWs (GWs). Notably, if the phase transition is of the first order, it imparts a distinctive oscillatory feature to the GW spectrum~\cite{An:2020fff,An:2022cce}. 
However, an intriguing aspect of our model is the predicted frequency range of these GWs. As our analysis will demonstrate, the frequency band of the GWs generated during this phase transition exceeds the current detection capabilities of observatories such as LIGO. This places the signals firmly beyond the reach of present-day GW detectors, which are optimized for lower frequency ranges.

This revelation points towards the necessity for future GW observatories with enhanced sensitivity to higher frequency bands. The detection of these high-frequency GWs would not only serve as a direct probe into the dynamics of early-universe phase transitions but also provide empirical evidence supporting the mechanisms of baryogenesis proposed in our model. Thus, our findings underscore the importance of advancing GW detection technology, paving the way for future detectors that can unlock these cosmic secrets and offer a deeper understanding of the universe's formative processes.

\begin{table}[htp]
\begin{center}
\begin{tabular}{c|ccc}
\hline
  ~&~ $\phi$ ~&~ $\chi$ ~&~ $\sigma$ \\
\hline
\hline
$U(1)$ ~&~ 0 ~&~ 1 ~&~ 0  \\
${\cal Z}_2$ ~&~ 1 ~&~ 1 ~&~ -1 \\  
\hline
\end{tabular}
\end{center}\caption{The charge assignment for the fields in this model.}
\label{tab:1}
\end{table}%

The rest of the paper is organized as follows. Section~\ref{sec:model} introduces the detailed framework of our model. Section~\ref{sec:genesis} is dedicated to the calculation of the $U(1)$ number generated prior to the phase transition. We delve into the dynamics of the inflaton field and its interaction with the complex scalar field $\chi$, which leads to the generation of the $U(1)$ charge.

Section~\ref{sec:convert} outlines the mechanism by which the $U(1)$ number produced during inflation is converted into the baryon number. This section also addresses the leading washout effects post-conversion and explores the conditions under which these effects are negligible. Section~\ref{sec:washout} focuses on estimating the washout effect induced by the phase transition. Such effect imposes constraints on the parameter space of our model, thereby limiting the observable signatures discussed subsequently. Section~\ref{sec:DM} investigates the production of dark matter as a resultant by-product of our model. Section~\ref{sec:observe} is devoted to the calculation of observable signals, including the cosmological collider physics signal and the GW signal. The methodology for calculating the cosmological collider signal follows closely with the approach presented in \cite{Bodas:2020yho}. For the sake of completeness, we include the key steps of this calculation in Appendix \ref{app:cc}. Finally, Section~\ref{sec:summary} offers a summary of our findings, encapsulating the key insights and implications of our study.

\section{The model }
\label{sec:model}

Given the radiative stability and the approximate shift symmetry of the inflaton sector, we can safely introduce derivative interactions involving the inflaton and other sectors. The field content of the model is discussed in the introduction section, and the fields together with their charges are listed in Table~\ref{tab:1}. Due to the approximate shift symmetry, it is tempting to define $\phi$ as a would-be goldstone boson corresponding to the breaking of some symmetry. Thus, the leading term of the broken current can be written as $F_\phi \partial_\mu \phi$, where $F_\phi$ is related to the symmetry breaking scale. Then the coupling between $\phi$ and $\chi$ can obtained by integrating out a heavy vector boson, 
\bea
{\cal L} = - \frac{g^2}{8 M^2} F_\phi \partial_\mu\phi i (\chi \partial^\mu \chi^* - \chi^* \partial^\mu \chi) \ .
\eea
From the effective theory point of view, this interaction can be rewritten as a dimension-five operator,
\bea\label{eq:lg1}
\ma{L}_{\text{dim-5}}= - \frac{i}{\Lambda} \partial_\mu\phi (\chi \partial^\mu \chi^* - \chi^* \partial^\mu \chi) \ ,
\eea
where 
\bea
\Lambda = \frac{8 M^2}{g^2 F_\phi} \ .
\eea
The evolution of $\phi_0$, the homogeneous component of $\phi$, is non-zero, and its time derivative gives rise to a {chemical potential}, $\mu = \dot\phi_0/\Lambda$, associated with the $U(1)$ number originating from the dim-5 operator. Here, the dot indicates differentiation with respect to physical time, $t$. In the usual inflation scenario, the velocity of the inflaton field can be tightly constrained as $\sqrt{\dot{\phi}}\gtrsim 60H$ based on CMB observations~\cite{Planck:2018jri}. Therefore, to ensure that the effective dim-5 operator remains valid during inflation, it is immediate to find that the cutoff $\Lambda$ must satisfy $\Lambda>60H$~\cite{Chen:2018xck,Wang:2019gbi,Wang:2020ioa,Bodas:2020yho}.

The observation that a non-zero chemical potential, $\mu \neq 0$, leads to violations of charge conjugation and combined charge-parity symmetries aligns with the prerequisites outlined by the Sakharov conditions for baryogenesis. This alignment naturally suggests the potential to interpret the $U(1)$ charge carried by the complex field $\chi$ as akin to the baryon number. The evolving inflaton field, in this context, offers a mechanism to materialize $\chi$ particles from the vacuum, ostensibly paving the way for baryogenesis. However, a critical examination reveals a significant hurdle: despite the presence of a non-zero $\mu$, the $U(1)$ number conservation remains intact. This conservation effectively precludes the realization of genuine baryogenesis within this framework, as the generation of baryon number necessitates the violation of its conservation.

To circumvent this challenge and facilitate the production of baryon number, our model introduces specific interactions that explicitly break the conservation of baryon number. The cornerstone of this approach is the explicit breaking term detailed in Eq.~(\ref{eq:Aterm}). This term is meticulously designed to disrupt the conservation of the $U(1)$ number, thereby laying the foundational mechanism for baryon number production within our theoretical construct. Through this strategic intervention, we aim to bridge the gap between the theoretical framework and the empirical phenomenon of baryogenesis, offering a viable pathway to generate baryon number in alignment with the conditions set forth by Sakharov.

Within our model, we posit that a critical event during the inflationary epoch is a symmetry restoration phase transition occurring within the $\sigma$ sector. Prior to the phase transition, the symmetry breaking mechanism allows $\sigma$ to acquire a non-zero vacuum expectation value (VEV), denoted as $\langle\sigma\rangle = v_\sigma$. This non-zero expectation value of $\sigma$ plays a pivotal role, as it activates a tadpole term for the $\chi$ field, emanating from the explicitly $U(1)$ symmetry-violating $A$-term~(\ref{eq:Aterm}). 

This tadpole term, in conjunction with the chemical potential $\mu$, acts as a catalyst for the generation of a $U(1)$ number density.

Once a symmetry-restoration phase transition occurs, the vacuum expectation value of $\sigma$ reverts to zero. This reversion has profound implications for the $U(1)$ number previously generated. Specifically, the $U(1)$ violating $A$-term, now in the absence of a non-zero $\langle\sigma\rangle$, begins to perturbatively wash out the $U(1)$ number.

Before the phase transition occurs, the dynamics of the $\chi$ sector are governed by a specific Lagrangian, which can be expressed as follows:
\begin{keyeqn}
\begin{align}\label{eq:lag}
\sqrt{-g}{\cal L} = a^3 |\dot\chi|^2 - a |\partial_i\chi|^2 + ia^3 \mu (\chi \dot\chi^* - \chi^* \dot\chi)- a^3 \left(m_\chi^2 |\chi|^2 -  A v_\sigma^2 (\chi + \chi^*) \right),
\end{align}
\end{keyeqn}
where $g$ denotes the determinant of the background metric, which adheres to the form $g_{\mu\nu}dx^{\mu}dx^{\nu}=-dt^2+a^2d\mathbf{x}^2$ with $\mu,\nu=0,1,2,3$. The scale factor $a$ is a critical component of the universe's expansion, and in the context of de Sitter inflation, it evolves exponentially with time as $a = e^{H t}$.

Within this framework, the baryon number associated with the $\chi$ field is defined by:
\begin{equation}\label{eq:current}
n_\chi = i (\chi^* \dot\chi - \chi \dot\chi^*) - 2 \mu|\chi|^2, \quad J_i = i (\chi \partial_i \chi^* - \chi^* \partial_i \chi).
\end{equation}
It is important to highlight that this definition of baryon number diverges from conventional definitions by incorporating an additional term that is proportional to the chemical potential $\mu$. This modification is a direct consequence of the inclusion of a dimension-five (dim-5) operator in our model. The significance of this term in the context of baryon number production will be elucidated further in our analysis.

From the Lagrangian provided above, we derive the conservation equation for $n_\chi$ and $J_i$:
\begin{equation}\label{eq:currentC}
\frac{d}{dt} (a^3 n_\chi) + \partial_i (a J_i) = - a^3 A v_\sigma^2 i (\chi^* - \chi).
\end{equation}
This equation reveals that due to the presence of the $U(1)$ symmetry-breaking term, the current associated with the baryon number is not conserved. Instead, its variation is directly proportional to a term that depends on both the coupling constant $A$ and the vacuum expectation value (VEV) of the $\sigma$ field. Notably, the influence of the $U(1)$ breaking term scales with $a^3$, indicating that in scenarios where $v_\sigma \neq 0$ and the imaginary part of $\chi$ is non-zero, the physical $U(1)$ number density, $n_\chi$, will not be diluted by the expansion of the universe. This analysis remains valid even if the chemical potential $\mu$ varies with time, as the derived conservation equation is a direct consequence of the equation of motion for the fields involved.

To deepen our understanding of baryogenesis within our theoretical framework and streamline the computations that follow, it is instructive to perform a $U(1)$ transformation specifically tailored for the $\chi$ sector. This transformation is given by:
\begin{equation}
\chi = \tilde\chi e^{- i \mu t},
\end{equation}
where $\tilde\chi$ represents the transformed field. This redefinition of the field allows us to recast the Lagrangian in a form that is more conducive to analyzing the dynamics of baryogenesis. Following this transformation, the Lagrangian takes the form:
\begin{keyeqn}
\begin{align}\label{eq:lag2}
\sqrt{-g}{\cal L} = a^3 \left[|\dot{\tilde\chi}|^2 - (m_\chi^2+\mu^2) |\tilde\chi|^2 \right] - a \partial_i \tilde\chi^* \partial_i \tilde\chi +  a^3 A v_\sigma^2 (\tilde\chi e^{- i \mu t} + \tilde\chi^* e^{i\mu t}).
\end{align}
\end{keyeqn}
This reformulated Lagrangian, denoted by Eq.~(\ref{eq:lag2}), reveals several key insights into the dynamics at play. Notably, the $U(1)$ transformation effectively serves as a form of diagonalization for the system. In scenarios where the $U(1)$ breaking term, specifically the $A$-term, is absent, the Lagrangian describes the dynamics of a free complex field, $\tilde\chi$, after undergoing the $U(1)$ transformation.

However, the presence of the $A$-term interaction within Eq.~(\ref{eq:lag2}) introduces explicit symmetry breaking, violating the charge conjugation, combined charge-parity, and $U(1)$ symmetries. This symmetry breaking is crucial for the mechanism of baryogenesis, as it provides the necessary conditions outlined by the Sakharov criteria for the generation of a baryon asymmetry in the universe.

Therefore, through this careful manipulation and analysis of the Lagrangian, we can conclude that the model possesses the requisite ingredients for successful baryogenesis. The explicit breaking of $C$, $CP$, and $U(1)$ symmetries, facilitated by the $A$-term interaction, sets the stage for the generation of a net baryon number, aligning with the overarching goals of our theoretical exploration.

\section{$U(1)$ number genesis before the phase transition}
\label{sec:genesis}

It is crucial to reiterate that the spectator field during inflation typically does not acquire a nonvanishing background. However, in our model, the spectator field $\chi$ does possess an inhomogeneous background due to the $A$-term. To illustrate this inhomogeneous background more clearly, we can derive the equation of motion for the homogeneous component of $\tilde{\chi}$ from \eqref{eq:lag2}:
\bea\label{eq:eom}
\ddot\tchi + 3H \dot\tchi +   (m_\chi^2 + \mu^2) \tchi =  Av_\sigma^2 e^{i\mu t} \ . 
\eea
We observe that the $A$-term appears on the right-hand side of the equation of motion, contributing to an inhomogeneous solution for $\tilde{\chi}$. What is more significant is that the source term $A v_\sigma^2 e^{i\mu t}$ is time-dependent, leading to a time-varying inhomogeneous solution:
\begin{equation}\label{eq:chisol}
\tchi=v_\chi e^{i(\mu t+\varphi)}+c_{\pm}e^{-\frac{3}{2}Ht\pm\sqrt{\frac{9}{4}H^2-(m_\chi^2+\mu^2)}}\ ,
\end{equation}
where
\bea\label{eq:vacuum}
~v_\chi = \frac{A v_\sigma^2}{(m_\chi^4 + 9 H^2 \mu^2)^{1/2}},~\sin\varphi = - \frac{3 H \mu}{(m_\chi^4 + 9 H^2 \mu^2)^{1/2}},
\eea
and $c_{\pm}$ represent the coefficients for the homogeneous solutions, which can be determined by the initial conditions. However, we will not analyze the homogeneous solutions in detail, as they are diluted by cosmic inflation. The radial component of the inhomogeneous solution is time-independent, with the time dependence manifesting in the angular component. Therefore, the expansion of our universe merely causes the inhomogeneous solution to rotate. It is worth noting that this unique time-dependent inhomogeneous solution enhances the cosmological collider signal, as discussed in \cite{Bodas:2020yho}.

Using the solution for $\tilde{\chi}$, we can perform an inverse $U(1)$ transformation to derive the expression for the original $\chi$ field. Substituting this expression for $\chi$ into the definition of the baryon current in \eqref{eq:current}, we can calculate the baryon number generated before the phase transition:
\begin{keyeqn}
\bea\label{eq:ini}
n_\chi^{(\rm {ini})} = - 2 \mu v_\chi^2 = - \frac{2 \mu A^2 v_\sigma^4}{m_\chi^4 + 9 H^2 \mu^2} \ .
\eea 
\end{keyeqn}
In summary, to generate a nonvanishing baryon number before the phase transition, three key ingredients are necessary. Firstly, a non-zero chemical potential for the spectator field carrying the baryon number is essential. Notably, the baryon number will approach zero in the limit as $\mu\to0$, despite the continued presence of a source term in the equation of motion \eqref{eq:eom}. Secondly, the inclusion of the $A$-term is crucial to provide a source term for the equation of motion for $\chi$. Without this term, only the homogeneous solution for $\chi$ remains, which is exponentially diluted in the late-time limit. Lastly, a $U(1)$ breaking term, controlled by the $\sigma$ sector in our model, is required.

The time scale of a first-order phase transition is determined by $\beta^{-1}$, where $\beta = - dS/dt$, the changing rate of the bounce action (See \cite{An:2020fff} and \cite{An:2022cce} on the detail description of first-order phase transition during inflation). For the phase transition to complete, it is required that $\beta \gg H$ (see e.g. \cite{Guth:1981uk} for discussions). Thus, as a leading order estimate, the change of $n_\chi$ during the phase transition can be estimated as 
\bea
\Delta n_\chi \approx 2 \beta^{-1} A v_\sigma^2 v_\chi \sin\varphi = \frac{6 \mu A^2 v_\sigma^4 H \beta^{-1}}{(m_\chi^4 + 9 H^2 \mu^2)} \ .
\eea
Then, we have
\bea
\frac{|\Delta n_\chi|}{n_\chi^{\rm {ini}}} \approx \frac{3 H }{\beta} \ .
\eea
Thus, as long as $\beta \gg 3 H $, we can neglect the change of $n_\chi$ during the phase transition and use $n_\chi^{\rm {ini}}$ as the initial condition of $n_\chi$ for later evolution after the phase transition.

The most important washout effect is that the $A$-term induces symmetry-breaking mass term for $\chi$ at the loop level through the diagram shown in Fig.~\ref{fig:washout}. The diagram is logarithmically divergent, and its size can be estimated as 
\bea
\Delta_\chi^2 \sim \frac{A^2}{16\pi^2} \ .
\eea
Then, the induced $U(1)$ number violating mass can be written as 
\bea
\Delta M_\chi \sim \frac{\Delta_\chi^2}{m_\chi} \ .
\eea
For this oscillation not to ruin the production and not to washout the $U(1)$ number, we have to require 
\bea\label{eq:condition3}
\Delta M_\chi \ll \min[\Gamma_\chi , H_\chi] \ ,
\eea
where $\Gamma_\chi$ is the decay rate of $\chi$, and $H_\chi \sim m_\chi^2/M_{\rm {pl}}$, is defined as the Hubble parameter when the temperature of the Universe drops below $m_\chi$. If the condition (\ref{eq:condition3}) is satisfied, most of the $U(1)$ number will be transferred to other particles (e.g. the decay products of $\chi$) before oscillating to anti-$U(1)$ number. 

The value of $\Gamma_\chi$ depends on the detailed model of how $\chi$ is coupled to the Standard Model particles. The detailed model will also induce new washout effects. Therefore, before computing any observation signal, we need to present the detailed model first.

\section{Transfer the $U(1)$ number to baryons and washout effects}
\label{sec:convert}
To account for the observed baryon-anti-baryon asymmetry in the present universe, it is required that the $\chi$ field interacts with the Standard Model sector. We shall henceforth equate the $U(1)$ quantum number with the $B - L$ number and propose the interaction between $\chi$ and the SM fields. Prior to detailing this interaction, it is essential to consider the implications of the sphaleron process. Two critical observations regarding the sphaleron process warrant attention \cite{Dine:2003ax}. First, if baryon and lepton numbers be independently generated in the early universe without an accompanying net $B - L$ number, the sphaleron process would act to annihilate both $B$ and $L$. Second, the presence of a net $B - L$ number would result in the sphaleron process preserving baryon and lepton numbers in magnitudes comparable to the initial $B - L$ value. Given the lack of constraints in such a high energy physics, there can be various ways to transfer the generated $U(1)$ number to the baryon number. In the following we only present a simple example. 

We introduce two additional colored scalar fields $\tilde X_1$ and $\tilde X_2$ to transfer the $U(1)$ number to SM particles. The quantum numbers of $\tilde X_1$ and $\tilde X_2$ are listed in Table.~\ref{tab:2}, and their interactions with $\chi$ and SM particles are
\bea
y_\chi \chi \tilde X_1 H^* \tilde X_2^* + y_1 \tilde X_1^*(Q e^c) + y_2 \tilde X_2 (d^c d^c)^* + {\rm {h.c.}} \ ,
\eea
where we can see that all the new interactions are marginal, and the decay rate of $\chi$ can be estimated as
\bea
\Gamma_\chi \approx \frac{y_\chi^2 m_\chi}{ 64\pi^3 } \ .
\eea
Here we assume that $m_\chi \gg m_{\tilde X_1} + m_{\tilde X_2}$. 

\begin{table}[htp]
\begin{center}
\begin{tabular}{c|cccc}
\hline
  ~&~ $SU(3)_C$ ~&~ $SU(2)_L$ ~&~ $U(1)_Y$~&~$U(1)_{B-L}$ \\
\hline
\hline
$\chi$          ~&~ 1 ~&~ 1 ~&~ 0 ~&~ -2 \\
$\tilde X_1$ ~&~ 3 ~&~ 2 ~&~ -7/6  ~&~ 4/3\\
$\tilde X_2$ ~&~ 3 ~&~ 1 ~&~ -2/3 ~&~ -2/3\\  
\hline
\end{tabular}
\end{center}\caption{The charge assignment for the fields in this model.}
\label{tab:2}
\end{table}%

From Table~\ref{tab:2} we can see that one unit of the $U(1)$ number defined in Table~\ref{tab:1} transfers to minus two units of $B-L$ number. The electroweak sphaleron effect forces $B + L = 0$. Thus, in the case that the washout effects are negligible, from Eq.~(\ref{eq:ini}), the physical baryon number density today can be written as 
\bea
n_B^{(0)} = \frac{2\mu A^2 v_\sigma^4}{m_\chi^4 + 9 H^2 \mu^2} z_{\rm {ph}}^{-3} \ ,
\eea
where $z_{\rm {ph}}$ measures the redshift from the phase transition to today. Thus, today's baryon-to-photon ratio can be calculated as 
\bea\label{eq:eta}
\eta = \frac{n_B^{(0)}}{n_\gamma} \approx 10^{-9} \times \left( \frac{H}{10^{14}~{\rm {GeV}}} \right)^{-1/2}\times \frac{c_A^2 c_\mu \theta}{9 c_\mu^2 + c_{m_\chi}^4}\times e^{- (3N_e - 29)} \ ,
\eea
where
\bea
c_A = \frac{A}{H} \ , \;\; c_\mu = \frac{\mu}{H} \ , \;\; c_{m_\chi} = \frac{m_\chi}{H} \ , \;\;  \theta = \frac{v_\sigma^4}{\rho_{\rm {inf}}} \ .
\eea

\section{The washout effects}\label{sec:washout}

In this model, there are two washout channels: (1) the $\chi \leftrightarrow\chi^*$ oscillation; (2) the $\chi-\sigma$ scatter.

\begin{figure}
 \centering
   \includegraphics[width=0.8\textwidth]{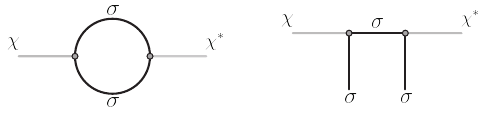}
\caption{The Feynman diagrams of washout processes.}
\label{fig:washout}
\end{figure}

\begin{itemize}

\item The $\chi-\chi^*$ oscillation process is induced by the diagram shown in Fig.~\ref{fig:washout}. The diagram is logarithmically divergent, and the result can be estimated as $\Delta m^2_\chi \approx A^2/(16\pi)^2$. Thus, the oscillation frequency can be estimated as $\Delta m_\chi^2/2m_{\chi}$. Therefore, for the oscillation not to significantly reduce the $U(1)$ number, we require the oscillation period is much shorter than the age of the universe when the temperature drops below $m_\chi$. This indicates 
\bea \label{eq:constraint21}
\frac{A^2}{ 32 \pi^2 m_\chi } = \frac{c_\chi m_\chi^2}{M_{\rm {pl}}} \ ,
\eea
with $c_\chi \ll 1$.

\item The Feynman diagram for $\chi-\sigma$ scattering process is shown in Fig.~\ref{fig:washout}, and the cross section can be estimated as 
\bea
\sigma_{\chi-\sigma} \approx \frac{A^4}{16\pi m_\sigma^6} \ .
\eea
During inflation after phase transition, the number density of $\sigma$ particle quickly decays. The initial number density of $\sigma$ particle is at most $\rho_{\rm {inf}}/m_\sigma$. Therefore, we require 
\bea
\frac{A^4}{16\pi m_\sigma^6} \frac{\rho_{\rm {inf}}}{m_\sigma} = c_{\chi\sigma} H_{\rm {inf}} \ ,
\eea
with $c_{\chi\sigma} \ll1$

\end{itemize}

From Table~\ref{tab:2} we can see that both $\tilde X_1$ and $\tilde X_2$ are charged under the SM gauge group and thus do not introduce new washout effects. 

Numerically, the requirement that $c_\chi \ll 1$ is always stronger than $c_{\chi\sigma} \ll1$. Thus, in the formula of $\eta$ we can use $c_\chi$ to replace $c_A$, and then Eq.~(\ref{eq:eta}) can be written as 
\bea\label{eq:eta2}
\eta\approx 8\times 10^{-10} \times \left(\frac{H}{10^{14}~{\rm{GeV}}}\right)^{1/2} \times \frac{c_\chi c_\mu c_{m_\chi}^3 \theta}{9 c_\mu^2 + c_{m_\chi}^4} \times e^{- 3(N_e -8)} \ .
\eea
In Eq.~(\ref{eq:eta2}), the combination $c_\mu c_{m_\chi}^3 / (9 c_\mu^2 + c_{m_\chi}^4) < 1/6$ due to the inequality of arithmetic and geometric means. The range of $\theta$ is determined by the shape of the potential of the $\sigma$ field, and cannot be fixed in the current study. $c_\chi$ is defined in Eq.~(\ref{eq:constraint21}) and should be smaller than one. One can see that if $\theta$ is not too large the symmetry restoration phase transition is likely to happen around $N_e$ equals eight e-folds before the end of inflation in the case of high scale inflation.

\section{Dark matter as a by-product}
\label{sec:DM}

As shown in Tables~\ref{tab:1} and \ref{tab:2}, in this simple model, $\sigma$ is the only field that is odd under the ${\cal Z}_2$ transformation. Dark matter production via symmetry-restoration phase transition has been discussed in \cite{An:2022toi}, in the case of high scale inflation, after the phase transition, the annihilation of the $\sigma$ particles provides negligible contribution to the number density of the $\sigma$ particle. Thus, after the phase transition, the physical number density of the $\sigma$ particle only dilutes with the expansion of the universe. However, in a realistic model, the phase transition sector may contain more fields. Thus, the lightest ${\cal Z}_2$ odd particle (LZP) can be a potential dark matter candidate. Assuming that $\sigma$ quickly decays into LZP after the phase transition. The LZP will first be relativistic, and its energy density redshift as $a^{-4}$. Then when the temperature in the $\sigma$ sector drops below $m_{\rm{LZP}}$, the LZP becomes non-relativistic and its energy density redshift as $a^{-3}$. Thus, the energy density of the LZP today can be estimated as 
\bea
L e^{- 3 N_e} \theta_{\rm{LZP}}\times \frac{g_{\star S}^{(0)} T_{\rm{CMB}}^3}{g_{\star S}^{\rm{RH}} T_{\rm RH}^3} \ ,
\eea
where $L$ is the latent heat density of the phase transition, and $\theta_{\rm{LZP}}$ accounts for the additional redshift of the relic energy density due to the transferrer from $\sigma$ to the LZP. In the case that the LZP is a direct decay product of $\sigma$, $\theta_{\rm{LZP}}$ can be estimated as $m_{\rm{LZP}}/m_\sigma$. However, since we have limited information in the $\sigma$ sector, we will not elaborate further model building.
Generically, the relic density of $\sigma$ can be written as
\bea
\Omega_{\rm{LZP}} = \theta_{\rm{LZP}}\frac{L e^{- 3 N_e}}{3 M_{\rm{pl}}^2 H_0^2} \times \frac{g_{\star S}^{(0)} T_{\rm{CMB}}^3}{g_{\star S}^{\rm{RH}} T_{\rm {RH}}^3} = \theta_L \theta_{\rm{LZP}}\left(\frac{H}{H_0}\right)^2  \times \frac{g_{\star S}^{(0)} T_{\rm{CMB}}^3}{g_{\star S}^{\rm{RH}} T_{\rm {RH}}^3} \ ,
\eea
where $H_0$ is today's Hubble expansion rate and $\theta_L$ is defined as $L/\rho_{\rm{inf}}$. Then, assuming instantaneous reheating we have 
\bea
\Omega_{\rm{LZP}} \approx 0.3 \times \frac{\theta_L \theta_{\rm{LZP}}}{2\times10^{-15}} \times\left( \frac{H}{10^{14}~{\rm{GeV}}} \right)^{1/2} e^{-3(N_e - 8)} \ .
\eea
Comparing to Eq.~(\ref{eq:eta2}) we can see that if we want to produce dark matter via the symmetry restoring phase transition, we should have $\theta_L \theta_{\rm{LZP}} \sim 10^{-15}$. In the case that $\theta_L \sim {\cal O}(1)$, we have $\theta_{\rm{LZP}}\sim 10^{-15}$. Thus, for high scale inflation, this indicates that the mass of LZP to be around TeV scale.

\section{Observational effects}
\label{sec:observe}

\subsection{The cosmological collider signal}\label{sec:ccsignal}

The {chemical potential}, $\mu$, and the explicit $U(1)$ number violating $A$-term in (\ref{eq:Aterm}) also induce an oscillatory non-gaussian correlation in the curvature perturbation. This signal will be observed by future cosmic microwave background (CMB) and large scale structure observations. The signal for the model we used in this work has been studied in \cite{Bodas:2020yho}, and here we list the result. For completion, we also put the main steps of the calculation in the Appendix \ref{app:cc}. The CMB can serve as a window to probe the physics at the energy scale of the inflationary Hubble constant, which is analogous to a cosmological collider \cite{ArkaniHamed2015, Shandera2010, Meerburg2010, Wang2020, Chen2010, Chen2018, Wang2020a, Cui2022, Lee2016, Chen2018a, Chen:2015lza, Lu:2019tjj, Liu:2019fag, Sou:2021juh, Bodas:2020yho}. During inflation, the spacetime horizon has a temperature $T = H/(2\pi)$, which allows for the production of particles with masses comparable to or smaller than the Hubble scale. These particles can interact with the inflaton field and leave distinctive imprints on its correlators, which are eventually observed in the CMB. By analyzing the CMB, we can extract information about the properties of these particles, such as their masses and spins, and thus explore the cosmological collider physics. However, the energy scale of the cosmological collider can be as high as $H \sim 10^{14} ~\text{GeV}$, which is far beyond the reach of the terrestrial colliders in the near future. To be specific, we are interested in the non-Gaussianity of the CMB, which can reveal the signatures of the cosmological collider physics. In our model, the inflaton is coupled with the massive $\chi$ field, which can consequently generate CC signal. Among the various measures of non-Gaussianity, the bispectrum is the most accessible one and will be the main focus of our analysis. 

\begin{figure}
 \centering
   \includegraphics[width=0.8\textwidth]{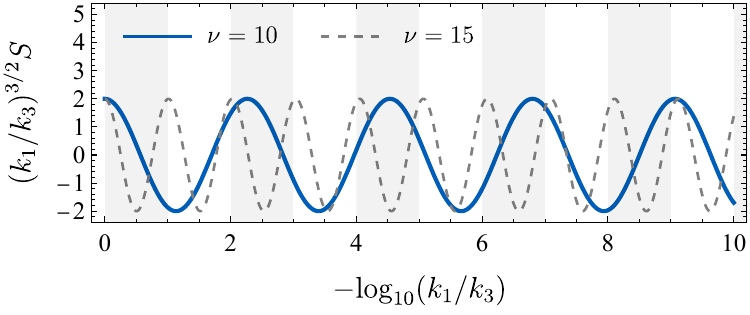}
\caption{The oscillatory feature of the shape function in the squeezed limit with $\mu/H = 40$ and $f_\text{NL}^{(\text{osc})} \simeq 1$.}
\label{fig:shapefunction}
\end{figure}

\begin{figure}[H]
 \centering
   \includegraphics[width=0.42\textwidth]{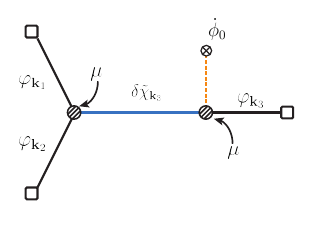}
   \includegraphics[width=0.42\textwidth]{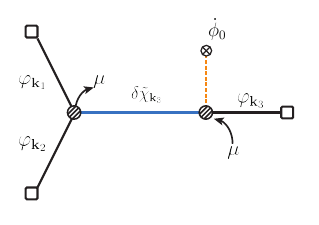}
\caption{The Feynman diagram corresponds to the process that generate the bispectrum computed in section \ref{sec:ccsignal}. Note that the energy is injected and removed at vertices due to the time-depend background $\tchi_0$.}
\label{fig:feynman}
\end{figure}

The bispectrum is captured by a dimensionless quantity, which is conventionally defined as 
\begin{align}
\qquad S(k_1, k_2, k_3)  = \fr{5}{6} \cdot \fr{\la \zeta_{\mb{k}_1} \zeta_{\mb{k}_2} \zeta_{\mb{k}_3} \ra'}{P_\zeta(k_1) P_\zeta(k_2) + P_\zeta(k_2) P_\zeta(k_3) + P_\zeta(k_3) P_\zeta(k_1) } \ , \label{eq:generalshapedef}
\end{align}
where $P_\zeta$ is the power spectrum and the prime on the correlator denotes the momentum conservation delta function is included implicitly.
This quantity encodes the shape of the bispectrum and is thus called the shape function. Typically, the CC signal appears in the squeezed limit, $k_3 \ll k_1 \sim k_2$, where the shape function can be approximated as 
\begin{keyeqn}
\begin{align}
S(k_1, k_2, k_3) \simeq |f^\text{(osc)}_\text{NL}| \left[\left(\fr{k_{1}}{k_3} \right)^{-3/2 + i(\nu - \mu)} + \text{c.c}  \right] \ , \label{eq:thedefofcc}
\end{align}
\end{keyeqn}
where $f^\text{(osc)}_\text{NL}$ is the amplitude of the CC signal and $\nu \equiv \sqrt{m_\text{eff}^2/H^2 - 9/4}$ with $m_\text{eff}^2 \equiv m_\chi^2 + \mu^2$. From \eqref{eq:thedefofcc}, we see an oscillatory dependence on the logarithm of the momentum ratio $k_1/k_3$ as shown in Fig \ref{fig:shapefunction}.
In particular, the frequency of such oscillation is $\nu$ which encodes the information of the mass of the spectator field. The shape function \eqref{eq:generalshapedef} can be calculated using the Schwinger-Keldysh formalism \cite{Chen2017}.  In the following, we consider the inflaton-spectator coupling that comes from the exponential terms in \eqref{eq:lag2}, which leads to the Feynman diagram in Fig.\ref{fig:feynman}. Note that the vertices in our Feynman diagram are time-dependent. To demonstrate this, we rewrite the equation of motion \eqref{eq:eom} of the redefined $\tchi$ field in terms of the conformal time, 
\bea \label{eq:eomconformal}
\left[ \partial_\eta^2 - \fr{2}{\eta}\partial_\eta + \left(-\nabla^2 + \fr{m_\text{eff}^2}{\eta^2} \right) \right] \tchi (\eta,\mb{x}) = A v_\sigma^2 \fr{(\eta/\eta_i)^{ i \mu }}{\eta^2}\ .
\eea
Here, $\eta_i$ depends on the initial value of the field and just appear as an unimportant overall phase.

The key observation is that the source term on the right hand side of \eqref{eq:eomconformal} leads to the non-trivial evolution of the homogeneous background $\tchi_0(\eta)$ of the $\tchi$ field, which have been separated as $\tchi = \tchi_0 + \delta \tchi$, where $\delta \tchi$ represents the fluctuation around the background. The equation of motion for $\tchi_0(\eta)$ is the following,
\bea
\left[ \partial_\eta^2 - \fr{2}{\eta}\partial_\eta + \fr{m_\text{eff}^2}{\eta^2}  \right] \tchi_0 (\eta) = A v_\sigma^2 \fr{(\eta/\eta_i)^{ i \mu }}{\eta^2} \ ,
\eea
which leads to the solution 
\bea\label{eq:eomtchibg}
\tchi_0(\eta) \simeq \fr{A v_\sigma^2 (-\eta)^{i \mu}e^{i\alpha}}{\sqrt{(m_\text{eff}^2 - \mu^2)^2 + 9\mu^2}} \equiv \tilde{\kappa}^{\dagger} (-\eta)^{i\mu} \ ,
\eea
where $\alpha$ is a constant phase. One direct consequence of \eqref{eq:eomtchibg} is that in the calculation of the inflaton correlators, there are now vertices that acquire a non-trivial time dependence $ (-\eta)^{i\mu}$ due to the chemical
potential. This effectively alters the energy of each vertex by an amount of order $\mu$,

\begin{figure}[H]
 \centering
   \includegraphics[width=0.49\textwidth]{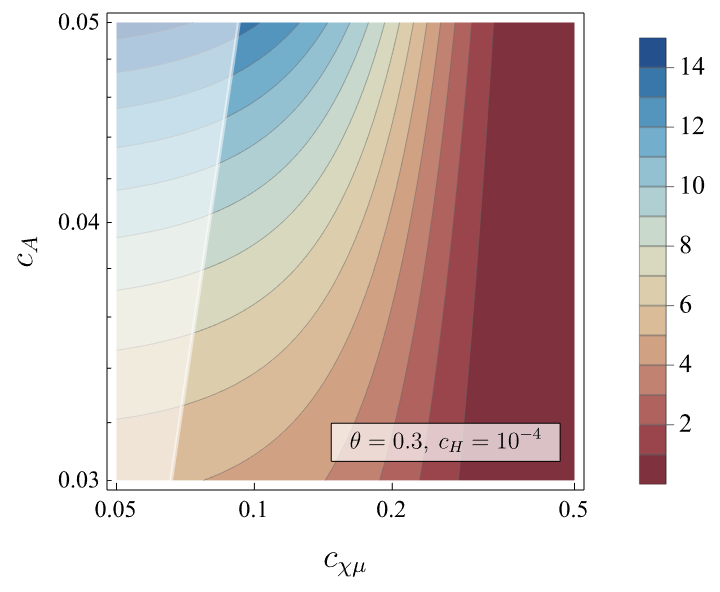}
   \includegraphics[width=0.49\textwidth]{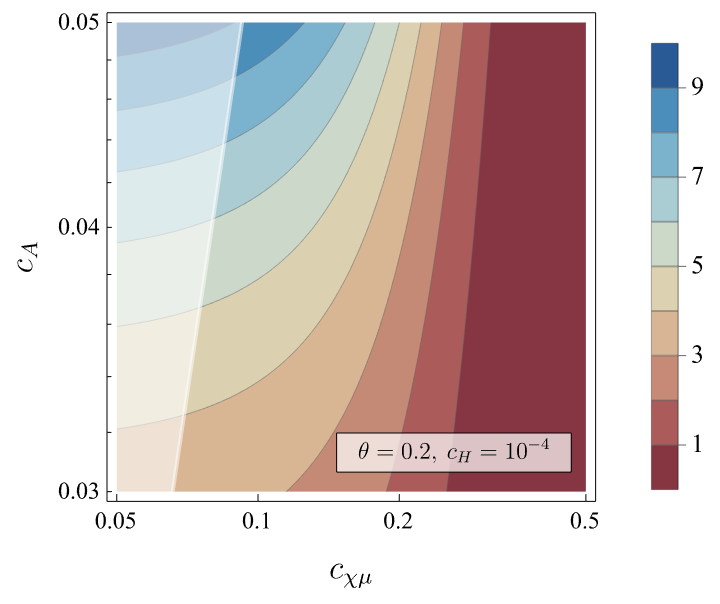}\\
   \includegraphics[width=0.49\textwidth]{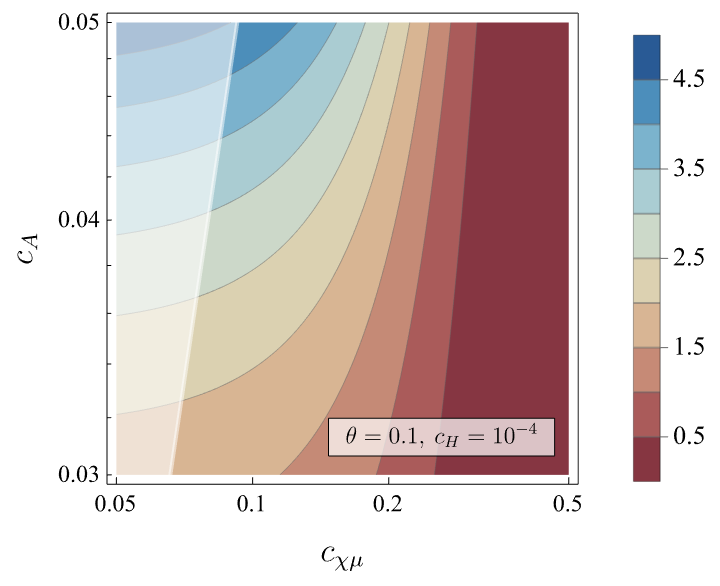}
   \includegraphics[width=0.49\textwidth]{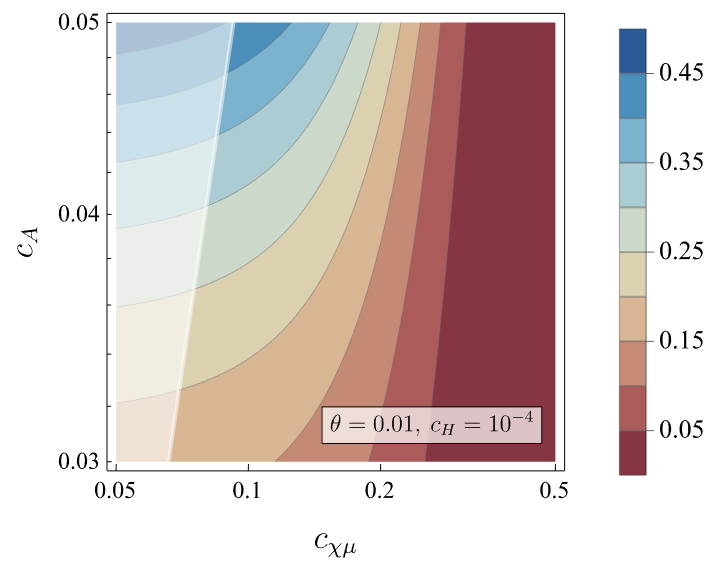}
\caption{The size of the cosmological collider signal $f_\text{NL}^{(\text{osc})}$ as a function of $c_A \equiv A/H$ and $c_{\chi\mu} \equiv m_\chi/\mu$ with  $c_H \equiv H / M_\text{Pl} = 10^{-4}$. The shaded region corresponds to the parameter space where $c_\chi \gtrsim 0.1$, which violates the washout constraint.}
\label{fig:fnlA}
\end{figure}
In Fig \ref{fig:fnlA}, we have shown $f^\text{(osc)}_\text{NL}$ as a function of $\chi-\sigma$ coupling strength $A$ and $c_{\chi\mu} \equiv m_\chi/\mu$ within the parameter space that is compatible with the constraint \eqref{eq:constraint21}. The detail of this derivation is demonstrated in Appendix \ref{app:cc}.

\subsection{GWs produced by the phase transition}

The GWs produced via bubble collisions during the phase transition might be observed today. Due to the interaction between the $\sigma$ and $\phi$, a backreaction from the $\sigma$ bubble collisions to the inflaton field will lead to the production of curvature perturbation, which will produce a secondary GW signal after the curvature perturbation reenters horizon after inflation. The curvature perturbation may also produce primordial black holes, which may become the dark matter candidate in today's universe. 

\begin{figure}
 \centering
   \includegraphics[width=0.49\textwidth]{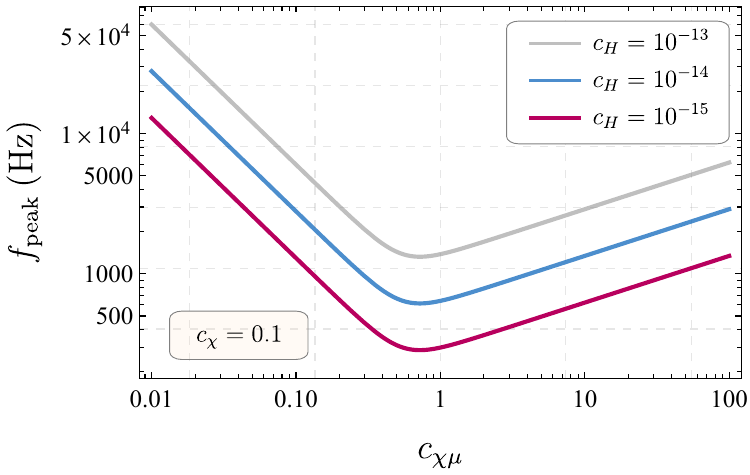}
   \includegraphics[width=0.49\textwidth]{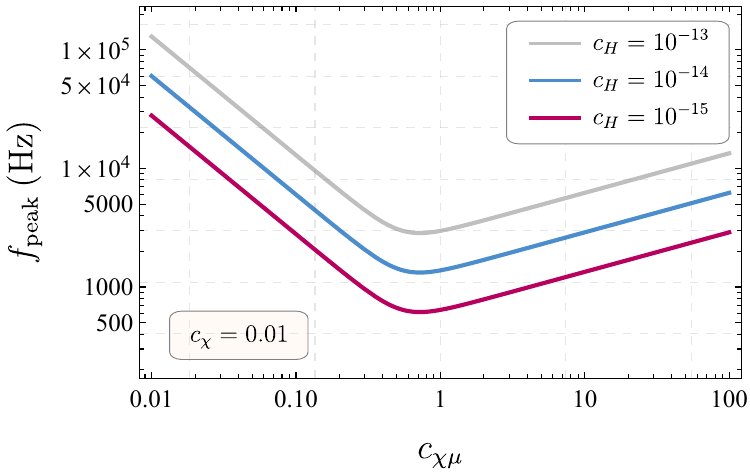}
\caption{An analysis of the peak frequency for GWs generated from symmetry restoration, depicted as a function of the mass of the $\chi$ particle and the ratio of chemical potential, across different inflationary Hubble scales.}
\label{fig:GWPeak}
\end{figure}

The strength and the shape of both the primary and secondary GWs are determined by the two parameters, $\beta$ and $L$. $\beta = - dS/dt$, is the changing rate of the bounce action during the phase transition. As shown in \cite{An:2020fff,An:2022cce,An:2023jxf}, the natural value for $\beta$ for a first-order phase transition induced by the evolution of the inflaton field is typically ${\cal O}(10)$. $L$ is the latent heat density of released during the phase transition. In \cite{An:2020fff,An:2022cce,An:2023jxf}, it is also shown that the shapes of the spectrums depend little on the details of the phase transition models. Today's spectra of the primary and secondary GWs can be written as 
\bea
\Omega^{(1)}_{\rm {GW}}(f) &=& \Omega_R \times \left(\frac{H}{\beta}\right)^5 \left( \frac{L}{\rho_{\rm {inf}}} \right)^2 {\cal F}_1 (\frac{f}{f_{\rm {peak}}}) \ ,\\
\Omega^{(2)}_{\rm {GW}}(f) &=& \frac{\Omega_R}{\epsilon^2} \times \left(\frac{H}{\beta}\right)^6 \left( \frac{L}{\rho_{\rm {inf}}} \right)^4 {\cal F}_2 (\frac{f}{f_{\rm {peak}}}) \ ,
\eea
where the $\Omega_R$ is the radiation energy density of the universe, $\epsilon$ is the slow-roll parameter after the phase transition, and ${\cal F}_{1,2}$ are the shape functions for the primary and secondary GWs, whose detailed forms can be found in~Refs.~\cite{An:2020fff,An:2022cce,An:2023jxf}. For $f$ around $f_{\rm {peak}}$, ${\cal F}_{1,2}$ are order one. The peak frequency is given by 
\bea
f_{\rm {peak}} = \alpha H \times (a_\star/a_{\rm {today}}) \ , 
\eea
where $\alpha$ is an order one parameter. The redshift factor counts $(a_\star/a_{\rm {today}})$ is composed by the redshift during inflation, $a_*/a_R$, and the redshifts are inflation, $a_R/a_0$. The former can be calculated using the conservation of the comoving baryon number after the symmetry restoration, that is
\be
n_\text{inf}^* a_*^3 = n_\text{inf}^R a_R^3 \ ,
\ee
where $n_\text{inf}^R$ is the baryon number at reheating. We can derive the redshift from the phase transition to the reheating through the following relation,
\begin{align}
\fr{a_*}{a_R} = & \left( \fr{n^R_\text{inf}}{n^*_\text{inf}} \right)^{1/3}= \left( \fr{n^R_\text{inf}/s_R}{n^*_\text{inf}/s_R} \right)^{1/3}= \left( \eta \times \fr{s_R}{n_\text{inf}^*} \right)^{1/3}, \label{eq:10}
\end{align}
where $\eta = 5\times 10^{-10}$ and $s_R$ is the entropy at the reheating. On the other hand, the reheating entropy can be derived through thermodynamics, and is given by,
\be
s_R = \fr{2 \pi^2 g^{(R)}_*  T_R^3}{45} \ , 
\ee
in which the reheating temperature is given by,
\be
T_R = \left[ \left(\fr{30}{g^{(R)}_* \pi^2}\right) 3 H^2 M_\text{Pl}^2 \right]^{1/4} \ .
\ee
Note that here we have assumed the scenario of instantaneous reheating.  The red-shift of the GW after the reheating can be calculated through the conservation of the comoving entropy density,
\be
s_R a_R^3 = s_0 a_0^3 \ ,
\ee
where $s_0$ and $a_0$ are the entropy and scaling factor of today's universe. Consequently, we have,
\be
 \fr{a_R}{a_0}  = \left( \fr{s_0}{s_R} \right)^{1/3} = \left(\fr{g^{(0)}_*}{g^{(R)}_*} \right)^{1/3} \fr{T_\text{CMB}}{T_R} \ ,
\ee
putting all the ingredients together, we have
\begin{align}
f_\text{peak} & = H \times \left( \eta \times \fr{2\pi^2 g_*^{(0)}}{45 \:n_{\text{inf}}} \right)^{1/3} \times T_{\text{CMB}} \nonumber \\
& = H \times  \left[ \eta \times \fr{g_*^{(0)} (c_{\chi\mu}^4 \mu^2 + 9 H^2 )M_\text{Pl}}{1440 c_\chi c_{\chi\mu}^3 \mu^2 v_\sigma^4 } \right]^{1/3} \times T_{\text{CMB}} \ ,
\end{align}
where $g^{(0)}_* \simeq 3.91$ is the today's value of the effective number of relativistic species. Therefore, the following expression can be derived,
\begin{keyeqn}
\begin{align}
f_\text{peak} & =  \left[ \fr{g_*^{(0)} (c_{\chi\mu}^4  + 9 c_{\mu}^{-2}  )c_H}{1440 c_\chi c_{\chi\mu}^3 \theta } \right]^{1/3} \times (\eta)^{1/3} \times T_{\text{CMB}} \ .
\end{align}
\end{keyeqn}
Note that there is a EFT bound $\mu \lesssim 60 H$. Therefore, we have the constraint $c_{\mu} < 60$. The peak frequency of the GW from the symmetry restoration is plotted in Fig \ref{fig:GWPeak}. 
Although the minimum frequency of the GW within the parameter space is beyond the reach of the current ground-based and space-based GW detector, it could be detected by the recent proposed new type of Fabry-Perot-Michelson interferometer \cite{Zhang:2022yab}.

\section{Summary and discussions}
\label{sec:summary}

In this work, we have demonstrated that during the inflationary period, a derivative interaction between the inflaton field and a scalar current, combined with an explicit symmetry-breaking term, can lead to the genesis of a $U(1)$ number. As detailed in Eqs.~(\ref{eq:vacuum}) and (\ref{eq:ini}), this $U(1)$ number is preserved within the vacuum configuration of the scalar field, $\chi$, and remains unaffected by the expansion of the universe. Subsequently, following a symmetry-restoration phase transition triggered by the evolution of the inflaton field, this $U(1)$ number is transferred to $\chi$ particles and begins to redshift along with the expansion of the universe. We have calculated the leading contributions to the washout effects and identified the parameter space where these effects are negligible.

As explored in Sec.~\ref{sec:DM}, the LZP can serve as a viable dark matter candidate and may be produced incidentally during the phase transition.

Further discussions in Sections \ref{sec:genesis} and \ref{sec:observe} highlight that both the chemical potential and the explicit symmetry-breaking term are essential for the processes of baryogenesis and for generating the CCP signal. Therefore, the CCP signal could serve as a definitive indicator of this baryogenesis scenario.

Additionally, we have investigated the GW signals associated with the phase transition. Unfortunately, to generate a sufficient baryon number, our model predicts a GW signal in the kHz range, which is currently beyond the detection capabilities of existing GW detectors. This limitation underscores the need for future advancements in GW detection technology to explore such high-frequency ranges.

\section*{Acknowledgement}

This work is supported in part by the National Key R\&D Program of China under Grant No. 2023YFA1607104 and 2021YFC2203100.

\appendix
\section{The estimation of the size of the cosmological collider signal}
\label{app:cc}
In this appendix, we present an analytical derivation of the cosmological collider signal strength as a function of the model parameters, most of the derivation in this section can be found in \cite{Bodas:2020yho}. We  expand the exponential coupling perturbatively in the order of $\Lambda$ as follows, 
\bea\label{eq:expexpand}
e^{- i \mu t - i \varphi /\Lambda} =  (- \eta)^{i \tilde{\mu} } \left( 1 - i \: \fr{\varphi}{\Lambda} - \fr{\varphi^2}{2\Lambda^2} + \cdots \right) \ .
\eea
Consequently, the leading Hamiltonian for the inflaton, $\tchi$ interaction reads
\begin{align}
\ma{H}_{\text{int}} = & \; g_1(-\eta)^{i \mu} \varphi \delta \tchi + g_2 (-\eta)^{i \mu} \varphi \delta \tchi \nonumber \\
& + g_3 (-\eta)^{i \mu} (\partial \varphi)^2 \delta \tchi + g_4 (-\eta)^{i \mu} \varphi^2 \delta \tchi +  \text{h.c} \ , \label{eq:hamiltonianapp}
\end{align}
where the coupling constants are given by $g_1 = 2 \mu \kappa $, $g_2 = i A v_\sigma^2/\Lambda$ $g_3 = -\kappa/\Lambda$ and $g_4 = A v_\sigma^2/(2\Lambda^2)$, with $\kappa \equiv -i A v_\sigma^2/ \sqrt{(m_\text{eff}^2 - \mu^2 )^2 + \mu^2}$.
In the following, we employ the in-in formalism to calculate the non-Gaussianity, which is given by,
\begin{align}
 \la\varphi_{\mb{k}_1} \varphi_{\mb{k}_2} \varphi_{\mb{k}_3} \ra'  = \la 0 |\; \bar{T} e^{i \int_{\eta_i}^{\eta_f} H_{\text{int}}^I \d \eta_1}  \varphi_{\mb{k}_1} \varphi_{\mb{k}_2} \varphi_{\mb{k}_3}  T e^{-i \int_{\eta_i}^{\eta_f} H_{\text{int}}^I \d \eta_2} \; | 0 \ra \ , \label{eq:inin}
\end{align}
where $T$, $\bar{T}$ denote the time ordering and the anti-time ordering operator respectively. Note that now there are two sets of vertices that come from both $\bar{T} e^{i \int_{\eta_i}^{\eta_f} H_{\text{int}}^I \d \eta_1}$ and $T e^{-i \int_{\eta_i}^{\eta_f} H_{\text{int}}^I \d \eta_2}$. Using the Hamiltonian given in \eqref{eq:hamiltonianapp}, the 3-point correlator of inflaton is given by the composition of the seed integrals,
\bea\label{eq:seedint}
\la\varphi_{\mb{k}_1} \varphi_{\mb{k}_2} \varphi_{\mb{k}_3} \ra' = \ma{I}_{+-} + \ma{I}_{+-} + \ma{I}_{++} + \ma{I}_{--} \ ,
\eea
where the seed integral is defined as the contributions from the leading terms of the expansion, i.e.
\begin{align}
\ma{I}_{+-} \equiv & \int_{-\infty}^{\eta_f} \fr{\d \eta_1}{\eta_1^4} \fr{\d \eta_2}{\eta_2^4} \cdot \left( \fr{\eta_2}{\eta_1}\right)^{i\mu} \cdot \la(g_1 \dot{\varphi} - g_2 \varphi) \delta\tchi^{\dagger} \cdot\varphi_{\mb{k}_1} \varphi_{\mb{k}_2} \varphi_{\mb{k}_3} \cdot (g_3 (\partial \varphi)^2 + g_4 \varphi^2) \delta\tchi \ra|_{\eta_f \to 0}\nonumber \\
&\qquad + (\mu \to -\mu) \ , \label{eq:ntoseedapp}\\
\ma{I}_{++} \equiv & \int_{-\infty}^{\eta_f} \fr{\d \eta_1}{\eta_1^4} \fr{\d \eta_2}{\eta_2^4} \cdot \left( \fr{\eta_2}{\eta_1}\right)^{i\mu} \cdot \la(g_1 \dot{\varphi} - g_2 \varphi) \delta\tchi^{\dagger} \cdot \varphi_{\mb{k}_1} \varphi_{\mb{k}_2} \varphi_{\mb{k}_3} \cdot (g_3 (\partial \varphi)^2 + g_4 \varphi^2) \delta\tchi \ra \nonumber \\
& \qquad + (\mu \to -\mu) \ . \label{eq:toseedapp}
\end{align}
The other two seed integrals can be obtained by taking the complex conjugation, $\ma{I}_{-+}=\ma{I}^*_{+-}$,$\ma{I}_{--}=\ma{I}^*_{++}$. By performing the Wick contraction, it is clear that the integrand of \eqref{eq:ntoseedapp} and \eqref{eq:toseedapp} are the product of the propagators which is given by the product of the mode function. 
For a given mode of the massive field fluctuation $\delta \tchi_\mb{k}$, it can be quantized as 
\bea
\delta \tchi_\mb{k} = a_{\mb{k}} u(k,\eta) + a_{-\mb{k}}^\dagger u^*(k,\eta) \ ,\label{eq:dtchiquantization}
\eea
and the mode function for the mass field reads
\bea
u(k,\eta) = \fr{\sqrt{\pi}}{2}e^{-\pi \nu/2} H \left(\fr{\eta}{k}\right)^{\fr{3}{2}} \text{H}^{(1)}_{i\nu} (\eta) \ ,
\eea
where  $\text{H}^{(1)}_{i\nu}$ denotes the Hankel function of first kind. Analogously, for the quantized massless inflaton field fluctuation, we have
\bea
 \varphi_\mb{k} = c_{\mb{k}} v(k,\eta) + c_{-\mb{k}}^\dagger v^*(k,\eta) \ ,
\eea
and the corresponding mode function is given by,
\bea
v(k,\eta) = \fr{H}{2 M_\text{Pl} \sqrt{\epsilon k^3}}(1 + i k \tau)e^{-i k \tau} \ . \label{eq:modefunctionofinflaton}
\eea
Plug \eqref{eq:dtchiquantization}-\eqref{eq:modefunctionofinflaton} into \eqref{eq:ntoseedapp} and \eqref{eq:toseedapp}, we can derive the analytical expression of the integrand. One would expect that in the end, the integral we have to deal with takes the following form,
\bea
\int_0^\infty \d x x^n e^{i \a x} \text{H}^{(\b)}_{i\nu}(x) \ , \label{eq:hankelintegrand}
\eea
where $\a = \pm$ and $\b = 1,2$. One can perform the integral in \eqref{eq:hankelintegrand} analytically. The time ordering $\theta$ function that arises from the Wick contraction in \eqref{eq:toseedapp} introduces a subtlety in the calculation of the seed integrals. There are two general methods to handle the $\theta$ function: the Melin-Barns transformation or the bootstrap method. We omit the details here, but we note that it is evident from the bootstrap equation that the $\theta$ function only affects the inhomogeneous term and does not alter the oscillatory feature of the seed integral or the CC signature. The dominant contribution to the 3-point correlator is given by
\bea
\la\varphi_{\mb{k}_1} \varphi_{\mb{k}_2} \varphi_{\mb{k}_3} \ra' \supset \ma{A}(\mu,\nu)\; r^{3/2 + i (\nu - \mu)} + \text{c.c} \ .
\eea
One can define a dimensionless shape function to characterize the bispectrum, which, in the squeezed limit, becomes
\begin{align}
S(k_1,k_2,k_3) \simeq & \fr{5}{12} \dot{\phi}_0 \fr{\la\varphi_{\mb{k}_1} \varphi_{\mb{k}_2} \varphi_{\mb{k}_3} \ra'}{\la\varphi_{\mb{k}_1} \varphi_{\mb{k}_1}  \ra'\la\varphi_{\mb{k}_3} \varphi_{-\mb{k}_3}  \ra'} \\
\simeq & \fr{5}{12} \dot{\phi}_0 \ma{A}(\mu,\nu)\; \left(\fr{k_{1}}{k_3} \right)^{-3/2 + i(\nu - \mu)} + \text{c.c} \ .
\end{align}
We introduce a dimensionless parameter $f_{\text{NL}}^{(\text{osc})} \equiv \fr{5}{12} \dot{\phi}_0 |\ma{A}|$ to quantify the magnitude of the CC signal.

\bibliography{refs}
\bibliographystyle{utphys}

\end{document}